\def\arrow{\rightarrow}
\def\threeeq{\equiv}        
\documentstyle[12pt,amssymbols]{article}
\def\etal{\hbox{\it et al.}}
\def\np#1#2#3{           {\it Nucl. Phys. }{\bf #1} (19#3) #2}
\def\pl#1#2#3{           {\it Phys. Lett. }{\bf #1} (19#3) #2}
\def\pr#1#2#3{           {\it Phys. Rev. }{\bf #1} (19#3) #2}

\def\prl#1#2#3{          {\it Phys. Rev. Lett. }{\bf #1} (19#3) #2}

\def\zp#1#2#3{           {\it Zeit. fur Physik }{\bf #1} (19#3) #2}
\def\blank#1#2#3{        {\bf #1} (19#3) #2}

\newfont{\bff}{cmtcsc10 scaled \magstep 1}
\begin{document}
\begin{titlepage}
\begin{center}
February 10, 1995     \hfill    LBL-35892 \\

\vskip .5in

{\large \bf Nonleptonic Two-Body Decays of $D$ Mesons in Broken
$SU(3)$}\footnote{This work was supported by the Director, Office of Energy
Research, Office of High Energy and Nuclear Physics, Division of High
Energy Physics of the U.S. Department of Energy under Contract
DE-AC03-76SF00098.}
\vskip .50in

Ian Hinchliffe and Thomas A.~Kaeding\footnote{Electronic addresses:
                                 {\tt theory@lbl.gov},
                                 {\tt kaedin@theorm.lbl.gov}.}

{\em Theoretical Physics Group\\
     Lawrence Berkeley Laboratory\\
     University of California\\
     Berkeley, California 94720}
\end{center}

\vskip .5in

\begin{abstract}

Decays of the $D$ mesons to two pseudoscalars, to two vectors, and
to pseudoscalar plus vector are discussed in the context of broken
flavor $SU(3)$.
A few assumptions are used to reduce the number of
parameters.
Amplitudes are fit to the available data, and
predictions of branching ratios for unmeasured modes are made.

\end{abstract}
\end{titlepage}

\renewcommand{\thepage}{\roman{page}}
\setcounter{page}{2}
\mbox{ }

\vskip 1in

\begin{center}
{\bf Disclaimer}
\end{center}

\vskip .2in

\begin{scriptsize}
\begin{quotation}
This document was prepared as an account of work sponsored by the United
States Government. While this document is believed to contain correct
information, neither the United States Government nor any agency
thereof, nor The Regents of the University of California, nor any of their
employees, makes any warranty, express or implied, or assumes any legal
liability or responsibility for the accuracy, completeness, or usefulness
of any information, apparatus, product, or process disclosed, or represents
that its use would not infringe privately owned rights.  Reference herein
to any specific commercial products process, or service by its trade name,
trademark, manufacturer, or otherwise, does not necessarily constitute or
imply its endorsement, recommendation, or favoring by the United States
Government or any agency thereof, or The Regents of the University of
California.  The views and opinions of authors expressed herein do not
necessarily state or reflect those of the United States Government or any
agency thereof, or The Regents of the University of California.
\end{quotation}
\end{scriptsize}

\vskip 2in

\begin{center}
\begin{small}
{\it Lawrence Berkeley Laboratory is an equal opportunity employer.}
\end{small}
\end{center}

\newpage
\renewcommand{\thepage}{\arabic{page}}
\setcounter{page}{1}

\section*{Introduction}

Many data are available on the hadronic two-body decays of charmed
mesons. Theoretical models that attempt to systematize the decay patterns have
been available for many years. These models usually make dynamical assumptions
in order to reduce the number of amplitudes that contribute to a particular
decay. For example, the large $N_c$ approximation
\cite{buras} \cite{oneovern},
or the heavy-quark effective theory \cite{bigi}.
It is not clear {\it a priori} how well such approximations should work
and hence how seriously to take a conflict between a prediction and a measured
value.
Another approach is to assume that the matrix elements factorize
\cite{werbol}. This  model is quite successful in describing observed modes,
but again, it is difficult to know whether a discrepancy is due to an incorrect
measurement of the failure of the assumption.
A more general approach based on a diagrammatic classification
\cite{CCH}, with different assumptions also exists.
In many cases attempts are made to obtain predictions of unmeasured modes
from these models.

$SU(3)$ is badly broken in these decays, so models based
on exact symmetry \cite{quigg} are not useful.
An attempt at a complete parameterization of the
$SU(3)$-breaking has been conspicuously missing,
due to the large number of reduced matrix elements involved.
We set out to remedy this omission.
This work gives a full parameterization of the
decays of the $D$ mesons into final states of two pseudoscalars (PP), two
vectors (VV) and a pseudoscalar plus a vector (PV), including $SU(3)$-breaking.
The elements of this parameterization--the particle representations,
the weak hamiltonian, the breaking operator, and the reduced matrix
elements--are discussed in the following sections.
We make only very few assumptions to limit the number of parameters.
We fit the parameters to the available data of two-body decays
and predict many unmeasured modes.
Because a few of the parameters are not constrained,
we indicate which branching fractions are needed to predict the rest
of certain classes of modes.
We comment on the case of $D_s \rightarrow \eta' \rho^+$,
where the model is barely consistent with data.

\section{Particle States in Flavor $SU(3)$}

In a model based on flavor $SU(3)$, the particles
are denoted by their $SU(3)$ representation.
The fundamental representation is the triplet ({\bf 3})
of quarks $u$, $d$, and $s$.
The three $D$ mesons \{$D^0$, $D^+$, $D_s^+$\} form an antitriplet
($\bar{\bf 3}$) representation.
The pseudoscalars \{$\pi^+$, $\pi^0$, $\pi^-$, $K^+$, $K^0$, $K^-$,
$\bar{K}^0$, $\eta_8$\}
form an octet ({\bf 8}) representation, as do the vectors
\{$\rho^+$, $\rho^0$, $\rho^-$, $K^{*+}$, $K^{*0}$,
$\bar{K}^{*-}$, $\bar{K}^{*0}$, $\omega_8$\}.
The $\eta_1$ and $\omega_1$ are each singlets.
The physical $\eta$, $\eta'$, $\phi$, and $\omega$ are linear combinations
of them, with mixing angles -17.3$^{\hbox{o}}$ \cite{CB} and 39$^{\hbox{o}}$
\cite{PDG94} for $\eta$-$\eta'$ and $\phi$-$\omega$
respectively.\footnote{$K^*$ denotes $K^*(892)$;  $\eta'$ denotes
$\eta'(958)$.}

\section{The Weak Hamiltonian}

The decays of the $D$ mesons are mediated by the weak hamiltonian.
Ignoring QCD corrections, the hamiltonian in terms of the quark fields is
\begin{equation}
\begin{array}{lcr}
H_{\hbox{\small weak}}
                 \; & = & \; \frac{G_F}{\sqrt{2}} cos^2 \theta_C
                          \; \bar{u} \gamma^{\mu} (1-\gamma_5) d
                          \; \bar{s} \gamma_{\mu} (1-\gamma_5) c \\
                    &   & + \; \frac{G_F}{\sqrt{2}} cos \theta_C sin \theta_C
                          \; \bar{u} \gamma^{\mu} (1-\gamma_5) s
                          \; \bar{s} \gamma_{\mu} (1-\gamma_5) c \\
                    &   & - \; \frac{G_F}{\sqrt{2}} cos \theta_C sin \theta_C
                          \; \bar{u} \gamma^{\mu} (1-\gamma_5) d
                          \; \bar{d} \gamma_{\mu} (1-\gamma_5) c \\
                    &   & - \; \frac{G_F}{\sqrt{2}} sin^2 \theta_C
                          \; \bar{u} \gamma^{\mu} (1-\gamma_5) s
                          \; \bar{d} \gamma_{\mu} (1-\gamma_5) c.
\end{array}
\label{hweakatMW}
\end{equation}
Note that the operators $\bar{q}$
create quarks and so transform as a triplet, while $q$ transforms as
the antitriplet.
Using the Clebsch-Gordan coefficients for the expansion of the product
{\bf 3} $\times$ $\bar{\bf 3}$ $\times$ {\bf 3}, we can
classify the operators according to irreducible representations of $SU(3)$
as follows:
\begin{equation}
\begin{array}{lcl}
(\bar{u}d)(\bar{s}c) & = & - \frac{1}{\sqrt{2}}
                           \bar{\bf 6} (\hbox{-}\frac{2}{3}, 1, 1)
                           - \frac{1}{\sqrt{2}}
                           {\bf 15} (\hbox{-}\frac{2}{3}, 1, 1), \\
(\bar{u}s)(\bar{d}c) & = &   \frac{1}{\sqrt{2}}
                           \bar{\bf 6} (\frac{4}{3}, 0, 0)
                           + \frac{1}{\sqrt{2}}
                           {\bf 15} (\frac{4}{3}, 1, 0), \\
(\bar{u}d)(\bar{d}c) & = &   \frac{1}{\sqrt{8}}
                           {\bf 3} (\frac{1}{3}, \frac{1}{2}, \frac{1}{2})
                           + \frac{1}{2}
                           {\bf 3'} (\frac{1}{3}, \frac{1}{2}, \frac{1}{2})
                           - \frac{1}{2}
                           \bar{\bf 6} (\frac{1}{3}, \frac{1}{2}, \frac{1}{2})
                           \\
                     &   &  - \frac{1}{\sqrt{3}}
                           {\bf 15} (\frac{1}{3}, \frac{3}{2}, \frac{1}{2})
                           - \frac{1}{\sqrt{24}}
                           {\bf 15} (\frac{1}{3}, \frac{1}{2}, \frac{1}{2}), \\
(\bar{u}s)(\bar{s}c) & = &   \frac{1}{\sqrt{8}}
                           {\bf 3} (\frac{1}{3}, \frac{1}{2}, \frac{1}{2})
                           + \frac{1}{2}
                           {\bf 3'} (\frac{1}{3}, \frac{1}{2}, \frac{1}{2})
                           + \frac{1}{2}
                           \bar{\bf 6} (\frac{1}{3}, \frac{1}{2}, \frac{1}{2})
                           + \sqrt{\frac{3}{8}}
                           {\bf 15} (\frac{1}{3}, \frac{1}{2}, \frac{1}{2}),
\end{array}
\label{quarkopsatMW}
\end{equation}
where $(\bar{q}q')$ denotes $\bar{q} \gamma^{\mu} (1 - \gamma_5) q'$.
The numbers in parentheses are hypercharge, total isospin, and third
component of isospin of the particular members of the $SU(3)$ representations.
The weak hamiltonian can now be written in terms of the
representations {\bf 3}, {\bf 3$'$}, $\bar{\bf 6}$, and {\bf 15} as
\begin{equation}
\begin{array}{lcl}
H_{\hbox{\small weak}} \;
             & = & G_F \sin^2 \theta_C
                   \left[ -\frac{1}{2} \bar{\bf 6}(\frac{4}{3}, 0, 0)
                   -\frac{1}{2} {\bf 15}(\frac{4}{3}, 1, 0) \right] \\
             & + & G_F \cos^2 \theta_C
                   \left[ -\frac{1}{2}
                   \bar{\bf 6}(\hbox{-}\frac{2}{3}, 1, 1)
                   -\frac{1}{2} {\bf 15}(\hbox{-}\frac{2}{3}, 1, 1) \right] \\
             & + & G_F \cos \theta_C \sin \theta_C \\
             &   & \left[ \frac{1}{\sqrt{2}}
                   \bar{\bf 6}(\frac{1}{3}, \frac{1}{2}, \frac{1}{2})
                   + \frac{1}{\sqrt{6}}
                   {\bf 15}(\frac{1}{3}, \frac{3}{2}, \frac{1}{2})
                   + \frac{1}{\sqrt{3}}
                   {\bf 15}(\frac{1}{3}, \frac{1}{2}, \frac{1}{2}) \right] .
\end{array}
\label{hopsatMWreps}
\end{equation}
Note that the {\bf 3} and {\bf 3$'$} representations
do not appear in the uncorrected $H_{\hbox{\small weak}}$ \cite{einhorn}.
Because the QCD corrections are multiplicative and do not mix the
$SU(3)$ representations, the {\bf 3} and {\bf 3$'$} will also not appear in
$H_{\hbox{\small weak}} (m_c)$.

Since the decays of the $D$ mesons occur at the scale of the
$c$-quark mass, we must allow the QCD evolution of the various operators
from the $W$-mass scale, where Equation (\ref{hweakatMW}) is valid, to the
$c$-mass scale.
The operators represented by the {\bf 15} are symmetric under quark
interchange, and those represented by the $\bar{\bf 6}$ are antisymmetric.
The QCD renormalization of operators with these symmetry
properties has been calculated \cite{gilman}.
We find that
\begin{equation}
\begin{array}{rlrl}
{\bf 15} & \arrow & {\bf 15} & \; \times \quad
     \left[ \frac{\alpha_s (M_W)}{\alpha_s (m_b)} \right]
           ^{{\Large {\it a}}^+_5} \quad \times \quad
     \left[ \frac{\alpha_s (m_b)}{\alpha_s (m_c)} \right]
           ^{{\Large {\it a}}^+_4}, \\
\bar{\bf 6} & \arrow & \bar{\bf 6} & \; \times \quad
     \left[ \frac{\alpha_s (M_W)}{\alpha_s (m_b)} \right]
           ^{{\Large {\it a}}^-_5} \quad \times \quad
     \left[ \frac{\alpha_s (m_b)}{\alpha_s (m_c)} \right]
           ^{{\Large {\it a}}^-_4}, \\
\end{array}
\label{otooprime}
\end{equation}
where
\begin{equation}
\begin{array}{lcc}
a^+_{\small{N_f}} & = & \frac{6}{33 - 2 N_f}, \\
a^-_{\small{N_f}} & = & \frac{-12}{33 - 2 N_f},
\end{array}
\label{anomalous}
\end{equation}
in the regime where there are $N_f$ flavor degrees of freedom.
Taking into account the change in the number of active flavors
as the $b$-quark threshold is crossed, and using $\alpha (M_Z) = 0.119$,
we obtain
\begin{equation}
\begin{array}{rclr}
    {\bf 15} & \arrow & 0.81 & {\bf 15}, \\
\bar{\bf 6}  & \arrow & 1.5  & \bar{\bf 6}. \\
\end{array}
\label{runningresults}
\end{equation}
With Equation (\ref{hopsatMWreps}) as the boundary condition, we have
\begin{equation}
\begin{array}{llll}
H_{\hbox{weak}}(m_c)
             & = &   & \frac{G_F}{2} \sin^2 \theta_C
                       \left[ -0.81 \; {\bf 15} (\frac{4}{3}, 1, 0)
                        -1.5 \; \bar{\bf 6} (\frac{4}{3}, 0, 0) \right] \\
             &   & + & \frac{G_F}{2} \cos^2 \theta_C
                       \left[ -0.81 \; {\bf 15} (\hbox{-}\frac{2}{3}, 1, 1)
                        -1.5 \; \bar{\bf 6} (\hbox{-}\frac{2}{3}, 1, 1)
                        \right] \\
             &   & + & \frac{G_F}{2} \cos \theta_C \sin \theta_C
                       \left[ 0.81 \times \frac{2}{\sqrt{3}} \;
                        {\bf 15} (\frac{1}{3}, \frac{1}{2}, \frac{1}{2})
                        \right. \\
             &   &   &  + \left. 0.81 \times \sqrt{\frac{2}{3}} \;
                          {\bf 15} (\frac{1}{3}, \frac{3}{2}, \frac{1}{2})
                        + 1.5 \times \sqrt{2} \;
                          \bar{\bf 6} (\frac{1}{3}, \frac{1}{2}, \frac{1}{2})
                          \right] .
\end{array}
\label{newhamiltonian}
\end{equation}
Note that the QCD corrections do not introduce any new phases into the process.
Unfortunately, until the values of the reduced matrix elements (discussed
below) are known, the coefficients in Equation (\ref{newhamiltonian})
are of little use.

\section{Parameterization}

\subsection{$SU(3)$ Breaking}

For a complete parameterization of any process in flavor $SU(3)$, we must
include explicit breaking.
Since we know that the source of flavor $SU(3)$ breaking
among the pions and kaons
is the difference between the quark masses,
we do this with an operator $M$ which transforms as an {\bf 8}.
Although the quark mass difference is insufficient to explain the large
$SU(3)$ breaking that will be found, an octet is the simplest
nontrivial operator that can be used.

We can express $M$ as
    \begin{equation}
    M = \alpha \lambda_3 + \beta \lambda_8,
    \label{middleM}
    \end{equation}
where $\lambda_i$ are the usual Gell-Mann matrices.
The term in $\alpha$ represents breaking of the isospin $SU(2)$ subgroup.
This breaking, proportional to the difference between up and down quark
masses, is expected to be very small and we neglect it in the following.
The constant $\beta$ can be absorbed into the reduced matrix elements.
Hence $M$ can be reduced to
    \begin{equation}
    M = \lambda_8.
    \label{newM}
    \end{equation}

\subsection{Reduced Matrix Elements}

Now consider the most general parameterization of the decays in the
context of the flavor $SU(3)$ symmetry.
For each possible contraction of the representations into an $SU(3)$
singlet there must be one parameter, {\it i.e.}, one reduced matrix element.
The representations involved are those in Section 1:  $D$ ($\bar{\bf 3}$),
$H$ ($\bar{\bf 6}$ + {\bf 15}), and two of P and V (each {\bf 1} or {\bf 8}).
In addition, we must include all possible ways of involving the
symmetry-breaking parameter $M$.
We assume that the breaking is linear in $M$.
Each reduced matrix element is, in principle, complex.
We have chosen to contract $D$ with $H$, then contract the products
(PP, PV, VV) (and then possibly with $M$), and finally contract the
two parts into the singlet.
Our labels for the reduced matrix elements reflects this.
For example, the matrix element denoted $(DH_{15})_8((\hbox{PP})_1M)_8$
is obtained by contracting $D$ and the {\bf 15} component of $H$ into
an octet, contracting PP into a singlet which combines with $M$ to
become another octet, and contracting the two resulting octets into
the singlet.

Unfortunately, the above parameterization involves far more parameters
than there exist data.
Therefore we make two important assumptions.
First, we assume that we can separate the spin and flavor dynamics
of the processes, {\it i.e.},
that the relative strengths of the reduced matrix
elements are the same in the PP, PV, and VV cases.
This implies that only forty-eight reduced $SU(3)$ matrix elements are needed.
They are labeled with $S$ and $O$ for the singlet and octet representations,
rather than with PP, PV, or VV.
In order to distinguish the spin states we introduce two parameters,
called ``PV'' and ``VV.''
Second, we assume that the phase of each reduced matrix element is given
solely by the representation of the product particles (before $M$ is included).
Bose symmetry for PP and VV
and an appropriate phase rotation
of the particle fields reduces the list of phases to
$(\eta_1 \eta_1)_1$, $(\eta_1 \omega_1)_1$, $(\omega_1 \omega_1)_1$,
(P$\eta_1$)$_8$, (P$\omega_1$)$_8$, (V$\eta_1$)$_8$, (V$\omega_1$)$_8$,
(PP)$_1$, (PP)$_{27}$,
(PV)$_1$, (PV)$_{8'}$, (PV)$_{10}$, (PV)$_{\bar{10}}$, (PV)$_{27}$,
(VV)$_1$, and (VV)$_{27}$.
One should note that we cannot determine the relative phases between
PP, PV, and VV.

The amplitude
for each decay mode can be expressed as a sum over the reduced matrix
elements with the appropriate Clebsch-Gordan coefficients:
  \begin{equation}
    A (D_j \rightarrow X_i) = \sum_k C_{ijk} R_k S_i.
  \label{sumofreduced}
  \end{equation}
Here $R_k$ are the reduced $SU(3)$ matrix elements and $S_i$ are the
parameters that we call PP $\threeeq$ 1, PV, and VV.
The $SU(3)$ Clebsch-Gordan factors $C_i$  were calculated by computer.
Many of the routines used are described in \cite{CPC}.

\subsection{Linear Combinations of Reduced Matrix Elements}

There are 45 measured values for the two-body decay modes and an additional
13 modes where upper limits exist.\footnote{The data are
from the Particle Data Group \cite{PDG94},
together with \cite{purohit} for the mode $D^+ \rightarrow K^{*0} \pi^+$.}
It would appear that there are still more parameters than data, and
therefore the model lacks predictability.
However, there are only forty linearly
independent combinations of the $SU(3)$ reduced matrix elements that
contribute to the possible decay modes of the $D$ mesons.
With the assumption of the last section concerning the phases of the reduced
matrix elements, the linear combinations fall into these classes:
  \vskip .25in
  \begin{center}
  \begin{tabular}{ll}
    involving $(SS)_1$:           & $L^{(1)}$                      \\
    involving $(SO)_8$:           & $L^{(2)}$, . . . , $L^{(8)}$   \\
    involving $(OO)_1$:           & $L^{(9)}$                      \\
    involving $(OO)_8$:           & $L^{(10)}$, . . . , $L^{(16)}$ \\
    involving $(OO)_{8'}$:        & $L^{(17)}$, . . . , $L^{(23)}$ \\
    involving $(OO)_{10}$:        & $L^{(24)}$, . . . , $L^{(28)}$ \\
    involving $(OO)_{\bar{10}}$:  & $L^{(29)}$, . . . , $L^{(33)}$ \\
    involving $(OO)_{27}$:        & $L^{(34)}$, . . . , $L^{(40)}$ \\
  \end{tabular}
  \end{center}
  \vskip .25in
We write them each as a sum over the reduced matrix elements, {\it viz.},
  \begin{equation}
    L^{(n)} = \sum_i C'_{in} R_i,
  \end{equation}
and normalize them for convenience by setting
  \begin{equation}
    \sum_i C_{in}^{\prime 2} = 1.
  \label{normL}
  \end{equation}
Now Equation (\ref{sumofreduced}) is replaced by
  \begin{equation}
    A (D_j \rightarrow X_i) = \sum_n C''_{ijn} L^{(n)} S_i.
  \label{sumofL}
  \end{equation}

The $L^{(n)}$ replace the reduced matrix elements in our parameterization
of the amplitudes.
The forty linearly independent combinations contain matrix elements
including those that involve the breaking operator $M$.
It is not possible to divide the linear combinations into a set that
contains only matrix elements without $M$ and a set containing only
matrix elements with $M$.
Of the forty combinations, three are not constrained by the available data.
We call them $L^{(1)}$, $L^{(2)}$, and $L^{(3)}$.
They are discussed below.

The replacement of the set of reduced matrix elements by the set of
linear combinations that contribute to the possible decay modes
reduces the number of parameters by eight.
The total number is now fifty three.
These parameters are fit to the data; the individual reduced matrix elements
are no longer considered.

The unconstrained combination $L^{(1)}$ contributes
to the modes
$D^0 \arrow \eta \eta$,
$\eta \eta'$, $\eta' \eta'$,
$\eta \phi$, $\eta \omega$, $\eta' \phi$, $\eta' \omega$,
$\phi \phi$, $\phi \omega$, and $\omega \omega$.
Because these modes are unobserved, the phases of
$(\eta_1 \omega_1)_1$, and $(\omega_1 \omega_1)_1$ are also unconstrained.
The remaining unconstrained linear combinations are $L^{(2)}$ and $L^{(3)}$.
They contribute to the above modes, and also to modes of the types
$D^0 \arrow \eta K^{0}$
and $D_s \arrow \eta K^+$.
By ``type'' we mean a class of modes that contain mesons of the
same flavors and charges.
Thus the type $D_s \arrow \eta K^+$ contains the modes
$D_s \arrow \eta K^+$, $\eta' K^+$, $\eta K^{*+}$, $\eta' K^{*+}$,
$\phi K^+$, $\omega K^+$, $\phi K^{*+}$, $\omega K^{*+}$, and no others.
With the exception of the limit
on the branching ratio for $D_s \arrow \phi K^+$,
there are no data for these modes.
We still have some freedom in the definition of $L^{(2)}$ and $L^{(3)}$
that allows modes of the type $D^0 \arrow \eta K^0$ to depend on only one
of them (choose $L^{(2)}$).
This will allow us to estimate one of their branching fractions and thereby
make some predictions of the other modes of this type.

\section{Data and Fitting Thereof}

The data used to determine the parameters are listed in Tables 1-5.
These are the modes for which there exist either experimental values or
experimental limits.
In the VV modes, S and D waves are possible.
Data exist from E691 \cite{e691} for the modes
$D^0 \rightarrow \bar{K}^{*0} \rho^0$ and
$D^+ \rightarrow \bar{K}^{*0} \rho^+$.
These are consistent with the S- and D-waves both having significant
amplitudes and are inconsistent with either being zero.
The ratios of S- and D-wave amplitudes
from these two modes are taken as additional data,
and the overall ratio of S- to D-wave amplitudes for the VV modes is
allowed to vary in the fit.
Its value is determined by the two modes mentioned above, and depends very
little on the other data.

For each mode we remove the phase space and Cabibbo factors and reduce
the branching ratio to a decay  amplitude in arbitrary units.
Because the vector particles have substantial widths, the phase space
for modes involving a vector is integrated over the relativistic
Breit-Wigner for that resonance.
The effect of this
is important for those modes where the sum of the particle masses
is within a few widths of the D-mass.
The modes $D^0 \arrow \phi K^{*0}$, $\phi \bar{K}^{*0}$, and
$D^+ \arrow \phi K^{*+}$ would be forbidden if the widths were set to zero.
Each amplitude is now expressed as a sum of Clebsch-Gordan coefficients
times the parameters that represent the reduced matrix elements,
and finally as a sum over the linearly independent combinations of
reduced matrix elements.

The parameters were fit to the data amplitudes with
{\bff MINUIT}, release 93.11 \cite{minuit}.
The total $\chi^2$ was found to be 30.9 for seven degrees of freedom,
indicating that the overall fit was poor.
However, more than half of the $\chi^2$ arose from only one mode.
The mode in question is $D_s \arrow \eta' \rho^+$.
The experimental value for the branching ratio
$D_s\to \eta^{\prime} \rho^+$ cannot be accommodated in our scheme.
It is measured \cite{cleo1} to be larger than that for $D_s\to \eta \rho^+$,
an {\it a priori} surprising result.
We note that the angular distribution of the decay pions
is barely consistent with that expected.
A confirmation of this experimental
value would be very significant as all other models \cite{werbol}
also predict a ratio of $B(D_s\to \eta^{\prime} \rho^+)/B(
D_s\to \eta \rho^+)$ of less than one.

We decided to reject the experimental value for the branching fraction
of $D_s \arrow \eta' \rho^+$.
The result is a better fit,
from which the branching ratios are reported in the tables.
The total $\chi^2$ is now 11.6 for six degrees of freedom.

\section{Predictions}

\subsection{Predictions from the Fit Parameters}

{}From the fit values of the parameters the branching ratios of decay modes
were calculated.
In Table 1 are presented the modes for which there exist
experimental values.
Our calculated branching ratios are consistent with the data,
with the exception of $D^+ \rightarrow \bar{K}^{*0} K^{*+}$
and $D_s \rightarrow \eta' \rho^+$.
For the former the fit prefers a branching ratio that is three standard
deviations below the reported experimental value.
The latter was removed before the fit (see Section 4) because its
experimental value was questioned.
For this mode we predict a branching ratio of (1.5 ${ }^{+1.9}_{-1.1}$)\%,
well below the reported experimental value \cite{cleo1}.
Tables 2-4 contain modes for which there is no experimental information
or for which there is an experimental limit.
We have attempted to predict the branching ratio of each mode from the fit.
However, in some cases the uncertainties are so large that we are able
only to provide (90\% confidence level) limits on the branching ratios.
Notice that in all cases in which there are experimental limits,
our predicted branching ratio or predicted limit is in the allowed region.
We are unable to say anything about the mode $D_s \rightarrow \rho^0 \pi^+$,
because the uncertainty on its prediction is greater than the
experimental limit.

There are two modes, $D_s \rightarrow \pi^+ \pi^0$ and
$D_s \rightarrow \rho^+ \rho^0$, which are forbidden in a model
without isospin breaking.
They are predicted to be identically zero.
The modes that are kinematically forbidden are $D^0 \rightarrow \eta' \eta'$,
$\eta' \phi$, and $\phi \phi$.
The modes involving the linear combinations $L^{(1)}$, $L^{(2)}$, and
$L^{(3)}$ are discussed below.
Any PP, PV, or VV mode not appearing in the tables is higher order in
the weak coupling $G_F$.

\subsection{Unconstrained Linear Combinations}

There remain three linearly independent combinations of the reduced
matrix elements that are not constrained by the data.
The combination $L^{(1)}$ contributes only to modes of the type
$D^0 \arrow \eta \eta$.
$L^{(2)}$ contributes to the types $D^0 \arrow \eta \eta$ and
$D^0 \arrow \eta K^0$.
$L^{(3)}$ contributes to these modes, and
to modes of the type $D_s \arrow \eta K^+$.

The first unconstrained linear combination $L^{(1)}$
contributes only to amplitudes involving $(SS)_1$.
These amplitudes, it is worth noting, are due entirely to $SU(3)$ breaking.
However, when we include the phases,
we must make four estimates in order to
obtain two predictions of modes of the type $D^0 \arrow \eta \eta$.
This would be an unproductive endeavor, and so we forego it.

In order to predict the modes of the types $D^0 \arrow \eta K^0$
and $D_s \rightarrow \eta K^+$, we need two new inputs.
In order to show the variability of the resulting predictions,
we try three different sets of inputs.
Scheme A is motivated by the recent CLEO measurement of the doubly-suppressed
mode $D^0 \rightarrow \pi^- K^+$ \cite{cleodoublesupp}, in which this mode
is found to have a branching ratio of about three times that of the
corresponding unsuppressed mode, $D^0 \rightarrow \pi^+ K^-$.
For this scheme, the two inputs are
  \begin{equation}
  \begin{array}{ll}
    B(D^0 \arrow \eta K^0) & =
        3 \tan^4 \theta_C \; B(D^0 \arrow \eta \bar{K}^0), \\
    B(D_s \arrow \eta K^+) & =
        3 \tan^2 \theta_C \; B(D_s \arrow \eta \pi^+). \\
  \end{array}
  \label{schemeaguess}
  \end{equation}
The linear combinations $L^{(2)}$ and $L^{(3)}$ are then constrained
and the remaining branching ratios in the column for scheme A in
Table 5 are found.
The predictions for scheme B are based on the following estimates:
  \begin{equation}
  \begin{array}{ll}
    B(D^0 \arrow \eta K^0) & =
        3 \tan^4 \theta_C \; B(D^0 \arrow \eta \bar{K}^0), \\
    B(D_s \arrow \phi K^+) & =
        3 \tan^2 \theta_C \; B(D_s \arrow \phi \pi^+). \\
  \end{array}
  \label{schemebguess}
  \end{equation}
A third scheme (C) is considered also.  It is based on these estimates:
  \begin{equation}
  \begin{array}{ll}
    B(D^0 \arrow \phi K^0) & =
        3 \tan^4 \theta_C \; B(D^0 \arrow \phi \bar{K}^0), \\
    B(D_s \arrow \phi K^+) & =
        \frac{1}{3} \tan^2 \theta_C \; B(D_s \arrow \phi \pi^+). \\
  \end{array}
  \label{schemecguess}
  \end{equation}
The resulting predictions are again in Table 5.
The spread in these values provides an indication of the expected ranges
for these quantities.

One should note that arbitrary choices of the above modes may fail to
give an acceptable fit, given the constraints from measured modes.
For example, an apparently reasonable choice would have been
  \begin{equation}
  \begin{array}{ll}
    B(D^0 \arrow \eta K^0) & =
          \tan^4 \theta_C \; B(D^0 \arrow \eta \bar{K}^0), \\
    B(D_s \arrow \eta K^+) & =
          \tan^2 \theta_C \; B(D_s \arrow \eta \pi^+). \\
  \end{array}
  \label{badguess}
  \end{equation}
A consistent fit cannot be obtained to implement this.
The parameters $L^{(2)}$ and $L^{(3)}$ could not be given values
to accommodate $B(D^0 \rightarrow \eta K^0) < 0.0052\%$.

\subsection{Modes Involving Axial Vectors}

There are a few modes involving axial vectors that have been observed or
for which there are experimental limits.
However, those that involve $K(1270)$ and $K(1400)$ are mixtures with
the $1^{+-}$ octet, which we can call B since it includes the $b_1(1235)$.
Therefore, in order to include these modes in our framework, we require
two new parameters, ``PA'' and ``PB.''
In addition, we must also accommodate the mixing between $f_{1(1)}$
and $f_{1(8)}$ to become $f_1(1285)$ and $f_1(1510)$, as well as the
new phases that are introduced.
There are too few experimental observations of the PA and PB modes to make
this endeavor fruitful.
For that reason, they are not included here.

\section{Comments on Models}

It is clear from the data alone that significant $SU(3)$ breaking is
necessary in any successful model of $D$ decays.
For example, $B(D^0 \arrow K^+ K^-)$ = $B(D^0 \arrow \pi^+ \pi^-)$
in exact $SU(3)$, yet they are in reality quite different.
Models based on exact $SU(3)$ \cite{quigg}, \cite{einhorn}, \cite{kingsley}
(or even on nonet symmetry \cite{rosen}) are thus not admitted by the data.

Models of $D$ decays based on heavy-quark effective theory
({\it e.g.}, \cite{bigi})
have as yet not developed to the point at which individual nonleptonic
decays can be calculated.
The question of whether HQET is applicable to the $c$ quark is still unsettled.
The HQET is based on an expansion in the parameter
  \begin{equation}
    \frac{\Lambda_{\hbox{\tiny QCD}}}{m_c} \simeq 0.2
  \end{equation}
and assumes that it is small.
Certainly this would be a good assumption in the case of the $b$ quark,
but perhaps not so here.
Until we are able to calculate branching fractions in HQET, we must
reserve judgement on its applicability to the $D$ mesons.

Diagrammatical methods to the problem of $D$ decays present us with
a complementary approach to the one adopted in this work.
The parameters in the $SU(3)$ framework represent sums of diagrams
in the diagrammatical approach.
A very general diagrammatical calculation of branching fractions appears
in \cite{CCH}.
Two shortcomings of their work lie in final-state interactions and in
the inclusion of $SU(3)$ breaking.
The phases of the final-state interactions are added to the model, and
are external to its central theme, and therefore appear as an ad-hoc
mechanism to force a fit.
$SU(3)$ breaking is added to the calculation as an additive correction
to the diagrams in which it is believed to be important.
However, there is also hidden breaking in the addition of phases in
the final-state interactions.
The result is a model in which the size and source of $SU(3)$ breaking
is not easily discerned.
It is difficult to draw any conclusions from the application of such a model.

The factorization method is a special case of the
diagrammatical approach.
In it certain diagrams are considered unimportant
({\it i.e.}, the annihilation diagrams).
However, \cite{werbol} find that these diagrams must be again included, as
well as final-state transitions and intermediate resonances.
The result is an eclectic
model with little elegance.  We are unable, because of the ad-hoc features,
to comment on the reliability and predictability of this model.

A description of nonleptonic $D$ decays in a large-$N_c$ (number of colors)
expansion \cite{oneovern} is an elegant one with few parameters.
In it, the source of $SU(3)$-breaking is introduced my including nearby
resonances.
It is also a subset of the diagrammatical approach and neglects some
diagrams based on their suppression by $1/N_c$.
One may argue that these diagrams are larger than thought, and cite
the fit of \cite{CCH} as evidence of this.
Nevertheless, \cite{oneovern} obtain excellent agreement with the data, with
the exception of some modes involving $\eta$ and $\eta'$.
In this model, $SU(3)$ breaking is introduced only through the inclusion
of resonances in one class of diagram.
They obtain, in agreement with our work, large breaking.

\section*{Conclusions}

There now exist enough data to constrain all but three combinations of the
reduced matrix elements of the broken $SU(3)$ model of the decays of $D$
mesons with the two assumptions discussed in Section 3.2.
We have used these data to do so.
Using the experimental information on 57 modes we are able to predict
branching ratios or upper limits for an additional 53 modes.
Only two measured modes are not easily accommodated in the fit.
The measurement of a few additional modes involving $\eta$, $\eta'$,
$\phi$, $\omega$ would enable another dozen or so modes to be predicted.

\section*{Acknowledgements}

Thanks go to G.\ Trilling for helping to reconcile the treatment of data
with experimental limits to that of data with positive measurements.

This work was supported by the Director, Office of Energy
Research, Office of High Energy and Nuclear Physics, Division of High
Energy Physics of the U.S. Department of Energy under Contract
DE-AC03-76SF00098.

\newpage

\begin{table}
\label{table:measured}
\caption{Modes with positive experimental values.
         Branching ratios from data and from the fit are given.}

\begin{center}
\begin{tabular}{|l|ll|ll|}
\hline
  Mode & \multicolumn{2}{l|}{Data BR} & \multicolumn{2}{l|}{Fit BR} \\
\hline
   $D^0$     $\rightarrow$           $K^-$                   $\pi^+$
    &    0.0401 & $\pm$    0.0014
    &    0.0400 & $\pm$    0.0014 \\
   $D^0$     $\rightarrow$           $K^-$                   $K^+$
    &    0.00454 & $\pm$    0.00029
    &    0.00453 & $\pm$    0.00030 \\
   $D^0$     $\rightarrow$           $\bar{K}^0$             $\pi^0$
    &    0.0205 & $\pm$    0.0026
    &    0.0208 & $\pm$    0.0022 \\
   $D^0$     $\rightarrow$           $\bar{K}^0$             $K^0$
    &    0.0011 & $\pm$    0.0004
    &    0.00103 & $\pm$    0.00043 \\
   $D^0$     $\rightarrow$           $\pi^-$                 $\pi^+$
    &    0.00159 & $\pm$    0.00012
    &    0.00159 & $\pm$    0.00012 \\
   $D^0$     $\rightarrow$           $\pi^-$                 $K^+$
    &    0.00031 & $\pm$    0.00014
    &    0.00031 & ${ }^{+   0.00018}_{-   0.00014}$ \\
   $D^0$     $\rightarrow$           $\pi^0$                 $\pi^0$
    &    0.00088 & $\pm$    0.00023
    &    0.00087 & $\pm$    0.00025 \\
   $D^0$     $\rightarrow$           $\eta$                  $\bar{K}^0$
    &    0.0068 & $\pm$    0.0011
    &    0.0069 & $\pm$    0.0011 \\
   $D^0$     $\rightarrow$            $\eta'$                $\bar{K}^0$
    &    0.0166 & $\pm$    0.0029
    &    0.0168 & $\pm$    0.0028 \\
   $D^0$     $\rightarrow$           $K^{*-}$                $\rho^+$
    &    0.059 & $\pm$    0.024
    &    0.063 & $\pm$    0.016 \\
   $D^0$     $\rightarrow$           $\bar{K}^{*0}$          $\rho^0$
    &    0.016 & $\pm$    0.004
    &    0.0164 & $\pm$    0.0038 \\
   $D^0$     $\rightarrow$           $\bar{K}^{*0}$          $K^{*0}$
    &    0.0029 & $\pm$    0.0015
    &    0.0029 & ${ }^{+   0.0019}_{-   0.0014}$ \\
   $D^0$     $\rightarrow$           $\omega$                $\bar{K}^{*0}$
    &    0.011 & $\pm$    0.005
    &    0.0099 & $\pm$    0.0044 \\
   $D^0$     $\rightarrow$           $\phi$                  $\rho^0$
    &    0.0019 & $\pm$    0.0005
    &    0.00192 & $\pm$    0.00045 \\
   $D^0$     $\rightarrow$           $K^-$                   $\rho^+$
    &    0.104 & $\pm$    0.013
    &    0.102 & $\pm$    0.013 \\
   $D^0$     $\rightarrow$           $K^-$                   $K^{*+}$
    &    0.0034 & $\pm$    0.0008
    &    0.00323 & $\pm$    0.00080 \\
   $D^0$     $\rightarrow$           $\bar{K}^0$             $\rho^0$
    &    0.0110 & $\pm$    0.0018
    &    0.0110 & $\pm$    0.0017 \\
   $D^0$     $\rightarrow$           $K^{*-}$                $\pi^+$
    &    0.049 & $\pm$    0.006
    &    0.049 & $\pm$    0.0058 \\
   $D^0$     $\rightarrow$           $K^{*-}$                $K^+$
    &    0.0018 & $\pm$    0.0010
    &    0.00209 & $\pm$    0.00087 \\
   $D^0$     $\rightarrow$           $\bar{K}^{*0}$          $\pi^0$
    &    0.030 & $\pm$    0.004
    &    0.0301 & $\pm$    0.0039 \\
   $D^0$     $\rightarrow$           $\phi$                  $\bar{K}^0$
    &    0.0083 & $\pm$    0.0012
    &    0.0081 & $\pm$    0.0012 \\
   $D^0$     $\rightarrow$           $\omega$                $\bar{K}^0$
    &    0.020 & $\pm$    0.004
    &    0.0195 & $\pm$    0.0043 \\
   $D^0$     $\rightarrow$           $\eta$                  $\bar{K}^{*0}$
    &    0.019 & $\pm$    0.005
    &    0.0204 & $\pm$    0.0049 \\
\hline
\end{tabular}
\end{center}
\end{table}

\begin{table}

\begin{center}
\begin{tabular}{|l|ll|ll|}
\hline
  Mode & \multicolumn{2}{l|}{Data BR} & \multicolumn{2}{l|}{Fit BR} \\
\hline
   $D^+$     $\rightarrow$           $\bar{K}^0$             $\pi^+$
    &    0.0274 & $\pm$    0.0029
    &    0.0262 & $\pm$    0.0028 \\
   $D^+$     $\rightarrow$           $\bar{K}^0$             $K^+$
    &    0.0078 & $\pm$    0.0017
    &    0.0086 & $\pm$    0.0016 \\
   $D^+$     $\rightarrow$           $\pi^0$                 $\pi^+$
    &    0.0025 & $\pm$    0.0007
    &    0.00257 & $\pm$    0.00067 \\
   $D^+$     $\rightarrow$           $\eta$                  $\pi^+$
    &    0.0075 & $\pm$    0.0025
    &    0.0068 & $\pm$    0.0021 \\
   $D^+$     $\rightarrow$           $\bar{K}^{*0}$          $\rho^+$
    &    0.021 & $\pm$    0.014
    &    0.0398 & $\pm$    0.0092 \\
   $D^+$     $\rightarrow$           $\bar{K}^{*0}$          $K^{*+}$
    &    0.026 & $\pm$    0.011
    &    0.0090 & ${ }^{+   0.0054}_{-   0.0041}$ \\
   $D^+$     $\rightarrow$           $\bar{K}^0$             $\rho^+$
    &    0.066 & $\pm$    0.025
    &    0.071 & $\pm$    0.018 \\
   $D^+$     $\rightarrow$           $\pi^+$                 $K^{*0}$
    &    0.00046 & $\pm$    0.00015
    &    0.00046 & $\pm$    0.00014 \\
   $D^+$     $\rightarrow$           $\bar{K}^{*0}$          $\pi^+$
    &    0.022 & $\pm$    0.004
    &    0.0217 & $\pm$    0.0041 \\
   $D^+$     $\rightarrow$           $\bar{K}^{*0}$          $K^+$
    &    0.0051 & $\pm$    0.0010
    &    0.00463 & $\pm$    0.00097 \\
   $D^+$     $\rightarrow$           $\phi$                  $\pi^+$
    &    0.0067 & $\pm$    0.0008
    &    0.00674 & $\pm$    0.00078 \\
   $D^+$     $\rightarrow$           $\phi$                  $K^+$
    &    0.00039 & $\pm$    0.00020
    &    0.00039 & ${ }^{+   0.00027}_{-   0.00020}$ \\
\hline
   $D_s$     $\rightarrow$           $\bar{K}^0$             $K^+$
    &    0.035 & $\pm$    0.007
    &    0.0319 & $\pm$    0.0059 \\
   $D_s$     $\rightarrow$           $\eta$                  $\pi^+$
    &    0.019 & $\pm$    0.004
    &    0.0204 & $\pm$    0.0039 \\
   $D_s$     $\rightarrow$            $\eta'$                $\pi^+$
    &    0.047 & $\pm$    0.014
    &    0.054 & $\pm$    0.012 \\
   $D_s$     $\rightarrow$           $\bar{K}^{*0}$          $K^{*+}$
    &    0.056 & $\pm$    0.021
    &    0.055 & $\pm$    0.018 \\
   $D_s$     $\rightarrow$           $\phi$                  $\rho^+$
    &    0.065 & $\pm$    0.017
    &    0.056 & $\pm$    0.014 \\
   $D_s$     $\rightarrow$           $\bar{K}^0$             $K^{*+}$
    &    0.042 & $\pm$    0.010
    &    0.043 & $\pm$    0.011 \\
   $D_s$     $\rightarrow$           $\bar{K}^{*0}$          $K^+$
    &    0.033 & $\pm$    0.005
    &    0.0328 & $\pm$    0.0053 \\
   $D_s$     $\rightarrow$           $\phi$                  $\pi^+$
    &    0.035 & $\pm$    0.004
    &    0.0349 & $\pm$    0.0040 \\
   $D_s$     $\rightarrow$           $\eta$                  $\rho^+$
    &    0.100 & $\pm$    0.022
    &    0.100 & $\pm$    0.019 \\
\hline
\end{tabular}
\end{center}

\end{table}

\begin{table}
\label{table:D0}
\caption{$D^0$ modes with predicted branching ratios.
         Experimental limits are given when available.
         All limits are at 90\% confidence.}

\begin{center}
\begin{tabular}{|l|l|ll|l|}
\hline
  Mode & Data BR & \multicolumn{2}{l|}{Predicted BR} & Predicted limit \\
\hline
   $D^0$     $\rightarrow$           $\pi^0$                 $K^0$
    &
    &    0.00017 & ${ }^{+0.00011}_{-0.00008}$ & \\
   $D^0$     $\rightarrow$           $\phi$                  $\bar{K}^{*0}$
    &
    &    0.00108 & ${ }^{+0.00073}_{-0.00054}$ & \\
   $D^0$     $\rightarrow$           $\rho^0$                $K^{*0}$
    &
    &    0.00038 & ${ }^{+0.00031}_{-0.00022}$ & \\
   $D^0$     $\rightarrow$           $\pi^0$                 $\rho^0$
    &
    &    0.00014 & ${ }^{+0.00018}_{-0.00011}$ & \\
   $D^0$     $\rightarrow$           $\pi^-$                 $\rho^+$
    &
    &    0.093 & ${ }^{+0.133}_{-0.075}$ & \\
   $D^0$     $\rightarrow$           $\rho^-$                $\pi^+$
    &
    &    0.094 & ${ }^{+0.136}_{-0.076}$ & \\
   $D^0$     $\rightarrow$           $\eta$                  $\pi^0$
    &
    &    0.0060 & ${ }^{+0.0092}_{-0.0050}$ & \\
   $D^0$     $\rightarrow$           $\eta$                  $\rho^0$
    &
    &    0.025 & ${ }^{+0.041}_{-0.021}$
      & $<$ 0.092 \\
   $D^0$     $\rightarrow$           $K^{*-}$                $K^{*+}$
    &
    &    0.0024 & ${ }^{+0.0041}_{-0.0021}$
      & $<$ 0.0092 \\
   $D^0$     $\rightarrow$           $\rho^0$                $K^0$
    &
    &    0.0024 & ${ }^{+0.0041}_{-0.0021}$
      & $<$ 0.0091 \\
   $D^0$     $\rightarrow$            $\eta'$                $\bar{K}^{*0}$
    & $<$   0.0011
    &    0.00018 & ${ }^{+0.00032}_{-0.00016}$
      & $<$ 0.00070 \\
   $D^0$     $\rightarrow$           $\rho^-$                $K^{*+}$
    &
    &    0.00022 & ${ }^{+0.00038}_{-0.00020}$
      & $<$ 0.00085 \\
   $D^0$     $\rightarrow$           $\rho^-$                $K^+$
    &
    &    0.0020 & ${ }^{+0.0035}_{-0.0018}$
      & $<$ 0.0078 \\
   $D^0$     $\rightarrow$           $\phi$                  $\pi^0$
    &
    &    0.024 & ${ }^{+0.049}_{-0.022}$
      & \\
   $D^0$     $\rightarrow$           $\pi^-$                 $K^{*+}$
    &
    &    0.0019 & ${ }^{+0.0037}_{-0.0018}$
      & $<$ 0.0080 \\
   $D^0$     $\rightarrow$           $\pi^0$                 $K^{*0}$
    &
    &    0.0020 & ${ }^{+0.0041}_{-0.0019}$
      & $<$ 0.0087 \\
   $D^0$     $\rightarrow$           $\bar{K}^0$             $K^{*0}$
    & $<$   0.0008
    & &
      & $<$ 0.00052 \\
   $D^0$     $\rightarrow$           $\omega$                $\pi^0$
    &
    & &
      & $<$ 0.086 \\
   $D^0$     $\rightarrow$            $\eta'$                $\rho^0$
    &
    & &
      & $<$ 0.011 \\
   $D^0$     $\rightarrow$           $\omega$                $\rho^0$
    &
    & &
      & $<$ 0.084 \\
   $D^0$     $\rightarrow$           $\rho^-$                $\rho^+$
    &
    & &
      & $<$ 0.015 \\
   $D^0$     $\rightarrow$           $\bar{K}^{*0}$          $K^0$
    & $<$   0.0015
    & &
      & $<$ 0.00061 \\
   $D^0$     $\rightarrow$            $\eta'$                $\pi^0$
    &
    & &
      & $<$ 0.057 \\
   $D^0$     $\rightarrow$           $\rho^0$                $\rho^0$
    &
    & &
      & $<$ 0.0065 \\
\hline
\end{tabular}
\end{center}

\end{table}

\begin{table}
\label{table:D+}
\caption{$D^+$ modes with predicted branching ratios.
         Experimental limits are given when available.
         All limits are at 90\% confidence.}

\begin{center}
\begin{tabular}{|l|l|ll|l|}
\hline
  Mode & Data BR & \multicolumn{2}{l|}{Predicted BR} & Predicted limit \\
\hline
   $D^+$     $\rightarrow$           $\rho^0$                $\rho^+$
    &
    &    0.0066 & $\pm$    0.0023 & \\
   $D^+$     $\rightarrow$           $\eta$                  $K^+$
    &
    &    0.0032 & ${ }^{+0.0030}_{-0.0020}$ & \\
   $D^+$     $\rightarrow$           $\pi^+$                 $K^0$
    &
    &    0.017 & ${ }^{+0.018}_{-0.011}$ & \\
   $D^+$     $\rightarrow$           $\pi^0$                 $K^+$
    &
    &    0.0086 & ${ }^{+0.0089}_{-0.0057}$ & \\
   $D^+$     $\rightarrow$           $\pi^0$                 $\rho^+$
    &
    &    0.0034 & ${ }^{+0.0036}_{-0.0023}$ & \\
   $D^+$     $\rightarrow$           $\phi$                  $K^{*+}$
    &
    &    0.00031 & ${ }^{+0.00035}_{-0.00022}$ & \\
   $D^+$     $\rightarrow$           $\rho^+$                $K^{*0}$
    &
    &    0.025 & ${ }^{+0.031}_{-0.018}$ & \\
   $D^+$     $\rightarrow$           $\rho^0$                $K^{*+}$
    &
    &    0.0095 & ${ }^{+0.0118}_{-0.0071}$ & \\
   $D^+$     $\rightarrow$           $\rho^+$                $K^0$
    &
    &    0.0087 & ${ }^{+0.0119}_{-0.0068}$ & \\
   $D^+$     $\rightarrow$           $\omega$                $K^{*+}$
    &
    &    0.0022 & ${ }^{+0.0031}_{-0.0018}$ & \\
   $D^+$     $\rightarrow$           $\omega$                $\pi^+$
    &    $<$ 0.007
    &    0.0024 & ${ }^{+0.0036}_{-0.0020}$ & \\
   $D^+$     $\rightarrow$           $\pi^0$                 $K^{*+}$
    &
    &    0.0103 & ${ }^{+0.0162}_{-0.0087}$ & \\
   $D^+$     $\rightarrow$           $\omega$                $\rho^+$
    &
    &    0.0026 & ${ }^{+0.0049}_{-0.0023}$
      & $<$ 0.011 \\
   $D^+$     $\rightarrow$           $\bar{K}^0$             $K^{*+}$
    &
    &    0.0012 & ${ }^{+0.0026}_{-0.0011}$
      & $<$ 0.0054 \\
   $D^+$     $\rightarrow$           $\eta$                  $\rho^+$
    & $<$   0.012
    &    0.0012 & ${ }^{+0.0022}_{-0.0011}$
      & $<$ 0.0048 \\
   $D^+$     $\rightarrow$            $\eta'$                $K^+$
    &
    &    0.0016 & ${ }^{+0.0041}_{-0.0015}$
      & $<$ 0.0082 \\
   $D^+$     $\rightarrow$           $\rho^0$                $K^+$
    &
    &    0.0018 & ${ }^{+0.0042}_{-0.0017}$
      & $<$ 0.0086 \\
   $D^+$     $\rightarrow$            $\eta'$                $\pi^+$
    & $<$   0.009
    &    0.00094 & ${ }^{+0.00237}_{-0.00092}$
      & $<$ 0.0048 \\
   $D^+$     $\rightarrow$           $\phi$                  $\rho^+$
    & $<$   0.015
    & &
      & $<$ 0.0074 \\
   $D^+$     $\rightarrow$           $\omega$                $K^+$
    &
    & &
      & $<$ 0.0012 \\
   $D^+$     $\rightarrow$            $\eta'$                $\rho^+$
    & $<$   0.015
    & &
      & $<$ 0.00071 \\
   $D^+$     $\rightarrow$           $\rho^0$                $\pi^+$
    & $<$   0.0014
    & &
      & $<$ 0.00091 \\
   $D^+$     $\rightarrow$            $\eta'$                $K^{*+}$
    &
    & &
      & $<$ 0.000082 \\
   $D^+$     $\rightarrow$           $\eta$                  $K^{*+}$
    &
    & &
      & $<$ 0.0022 \\
\hline
\end{tabular}
\end{center}

\end{table}

\begin{table}
\label{table:Ds}
\caption{$D_s$ modes with predicted branching ratios.
         Experimental limits are given when available.
         All limits are at 90\% confidence.}

\begin{center}
\begin{tabular}{|l|l|ll|l|}
\hline
  Mode & Data BR & \multicolumn{2}{l|}{Predicted BR} & Predicted limit \\
\hline
   $D_s$     $\rightarrow$           $\pi^0$                 $K^+$
    &
    &    0.0059 & ${ }^{+0.0048}_{-0.0034}$ & \\
   $D_s$     $\rightarrow$           $\pi^+$                 $K^{*0}$
    &
    &    0.038 & ${ }^{+0.047}_{-0.028}$ & \\
   $D_s$     $\rightarrow$            $\eta'$                $\rho^+$
    &          12.0 $\pm$ 3.0$^\dagger$
    &    0.015 & ${ }^{+0.019}_{-0.011}$ & \\
   $D_s$     $\rightarrow$           $\pi^0$                 $K^{*+}$
    &
    &    0.077 & ${ }^{+0.096}_{-0.058}$ & \\
   $D_s$     $\rightarrow$           $\rho^0$                $K^{*+}$
    &
    &    0.0126 & ${ }^{+0.0164}_{-0.0096}$ & \\
   $D_s$     $\rightarrow$           $\rho^+$                $K^0$
    &
    &    0.031 & ${ }^{+0.043}_{-0.024}$ & \\
   $D_s$     $\rightarrow$           $\rho^0$                $K^+$
    &
    &    0.049 & ${ }^{+0.071}_{-0.040}$ & \\
   $D_s$     $\rightarrow$           $\omega$                $\rho^+$
    &
    &    0.012 & ${ }^{+0.030}_{-0.012}$
      & $<$ 0.061 \\
   $D_s$     $\rightarrow$           $\pi^+$                 $K^0$
    & $<$   0.007
    &   &
      & $<$ 0.0015 \\
   $D_s$     $\rightarrow$           $K^0$                   $K^{*+}$
    &
    &   &
      & $<$ 0.00039 \\
   $D_s$     $\rightarrow$           $K^0$                   $K^+$
    &
    &  &
      & $<$ 0.00046 \\
   $D_s$     $\rightarrow$           $K^{*0}$                $K^{*+}$
    &
    &  &
      & $<$ 0.00057 \\
   $D_s$     $\rightarrow$           $\rho^+$                $K^{*0}$
    &
    &  &
      & $<$ 0.0080 \\
   $D_s$     $\rightarrow$           $K^{*0}$                $K^+$
    &
    &  &
      & $<$ 0.00025 \\
   $D_s$     $\rightarrow$           $\omega$                $\pi^+$
    & $<$   0.017
    &  &
      & $<$ 0.0090 \\
   $D_s$     $\rightarrow$           $\pi^0$                 $\rho^+$
    &
    &  &
      & $<$ 0.064 \\
   $D_s$     $\rightarrow$           $\rho^0$                $\pi^+$
    & $<$   0.0028
    &  &
      & not significant \\
    $D_s$ $\rightarrow$ $\pi^+$ $\pi^0$ & & $\equiv$ 0 & & \\
    $D_s$ $\rightarrow$ $\rho^+$ $\rho^0$ & & $\equiv$ 0 & & \\
\hline
\end{tabular}
\end{center}

\begin{tabular}{l}
\quad \quad \quad \quad \quad \quad \quad \quad \quad \quad \quad \quad \\
\cline{1-1}
\end{tabular}

{\small $^\dagger$See text.}

\end{table}

\begin{table}
\label{table:estimates}
\caption{Modes based on estimates.
         The only available experimental limit is shown.
         Values marked with * are inputs.}

\begin{center}
\begin{tabular}{|l|l|l|l|l|}
\hline
  Mode & Data BR & Fit BR & Fit BR & Fit BR\\
       &         & (scheme A) & (scheme B) & (scheme C) \\
\hline
   $D^0$     $\rightarrow$           $\eta$                  $K^0$
    &
    &    0.000054* & 0.000054* & 0.00035\\
   $D^0$     $\rightarrow$            $\eta'$                $K^0$
    &
    &    0.00046 & 0.00046 & 0.00085\\
   $D^0$     $\rightarrow$           $\phi$                  $K^{*0}$
    &
    &    0.000019 & 0.000019 & 0.000016\\
   $D^0$     $\rightarrow$           $\omega$                $K^{*0}$
    &
    &    0.00027 & 0.00027 & 0.0012\\
   $D^0$     $\rightarrow$           $\phi$                  $K^0$
    &
    &    0.00054 & 0.00054 & 0.000066* \\
   $D^0$     $\rightarrow$           $\omega$                $K^0$
    &
    &    0.000096 & 0.000096 & 0.0014\\
   $D^0$     $\rightarrow$           $\eta$                  $K^{*0}$
    &
    &    0.00048 & 0.00048 & 0.00094 \\
   $D^0$     $\rightarrow$            $\eta'$                $K^{*0}$
    &
    &    0.0000083 & 0.0000083 & 0.0000024 \\
\hline
   $D_s$     $\rightarrow$           $\eta$                  $K^+$
    &
    &    0.0027* & 0.00041 & 0.0031 \\
   $D_s$     $\rightarrow$            $\eta'$                $K^+$
    &
    &    0.017 & 0.052 & 0.015 \\
   $D_s$     $\rightarrow$           $\phi$                  $K^{*+}$
    &
    &    0.011 & 0.024 & 0.0095 \\
   $D_s$     $\rightarrow$           $\omega$                $K^{*+}$
    &
    &    0.0057 & 0.028 & 0.0046 \\
   $D_s$     $\rightarrow$           $\phi$                  $K^+$
    &  $<$  0.0025
    &    0.00051 & 0.0033* & 0.00037* \\
   $D_s$     $\rightarrow$           $\omega$                $K^+$
    &
    &    0.0064 & 0.019 & 0.0055 \\
   $D_s$     $\rightarrow$           $\eta$                  $K^{*+}$
    &
    &    0.00083 & 0.00015 & 0.00094 \\
   $D_s$     $\rightarrow$            $\eta'$                $K^{*+}$
    &
    &    0.00090 & 0.0028 & 0.00077 \\
\hline
\end{tabular}
\end{center}
\end{table}

\end{document}